\documentclass[journal]{IEEEtran}

\usepackage{pgfplots}
\usepackage[bookmarks,colorlinks]{hyperref}
\usepackage[linesnumbered,ruled,lined]{algorithm2e}
\usepackage{enumitem}
\usepackage{algpseudocode}
\usetikzlibrary{shapes.multipart,intersections}
\usepackage{cite}
\usepackage{amsmath,amssymb,amsfonts,amsthm,steinmetz}
\usepackage{graphicx}
\usepackage{mathrsfs}  
\usepackage{textcomp}
\usepackage{acronym}
\usepackage{xcolor}
\usepackage{upgreek,xspace}
\usepackage{array}
\usepackage{tikz}
\DeclareMathOperator{\atanh}{atanh}

\usepackage[all=normal,paragraphs=normal,floats=normal,mathspacing=normal,wordspacing=normal,charwidths=normal,mathdisplays=normal,leading=tight]{savetrees}

\usetikzlibrary{calc}
\makeatletter
\newcommand{\gettikzxy}[3]{%
  \tikz@scan@one@point\pgfutil@firstofone#1\relax
  \edef#2{\the\pgf@x}%
  \edef#3{\the\pgf@y}%
}
\makeatother

\usepackage[draft]{todonotes}

\usepackage{esvect}

\usepackage{times}
\usepackage{bm}
\usepackage{amsmath}
\usepackage{amssymb}
\usepackage{stmaryrd}
\usepackage{babel}
\usepackage{graphics, graphicx}
\usepackage{xcolor}
\usepackage{gensymb}
\usepackage{cite}
\usepackage{enumitem}
\usepackage{url}

\begin{document}

\title{Statistical Multiport-Network Modeling and \\Efficient Discrete Optimization of RIS}

\author{Cheima~Hammami,~Luc~Le~Magoarou,~\IEEEmembership{Member,~IEEE},~and~Philipp~del~Hougne,~\IEEEmembership{Member,~IEEE}
\thanks{
C.~Hammami and P.~del~Hougne are with Univ Rennes, CNRS, IETR - UMR 6164, F-35000, Rennes, France (e-mail: \{cheima.hammami; philipp.del-hougne\}@univ-rennes.fr).
}
\thanks{
L.~Le~Magoarou is with INSA Rennes, CNRS, IETR - UMR 6164, F-35000, Rennes, France (e-mail: luc.le-magoarou@insa-rennes.fr).
}
\thanks{\textit{(Corresponding Author: Philipp del Hougne.)}}
\thanks{This work was supported in part by the ANR France 2030 program (ANR-22-PEFT-0005) and the ANR PRCI program (ANR-22-CE93-0010).}
}

\maketitle

\begin{abstract}
This Letter addresses the physics-consistent optimization of reconfigurable intelligent surfaces (RISs) with mutual coupling (MC) and 1-bit-programmable RIS elements. This combination of constraints is typical of current prototypes but unexplored in theoretical work. \textit{First}, we present a simple statistical generator for multiport-network-theory (MNT) parameters of rich-scattering, RIS-parametrized channels. We account for reciprocity, passivity, and coherent backscattering; then, we add a simple hyper-parameter to control the MC strength. \textit{Second}, we benchmark model-agnostic (dictionary search, coordinate descent, genetic algorithm) and model-based (temperature-annealed back-propagation) strategies under varying MC, with and without intelligent initialization. Except when MC is negligible, coordinate descent with random initialization offers the best trade-off in performance, runtime, and memory. Our insights can guide wireless practitioners who optimize RIS prototypes and other reconfigurable wave systems.
\end{abstract}


\section{Introduction}

Reconfigurable intelligent surfaces (RISs) are emerging as an enabling technology to dynamically tailor wireless channels to diverse application needs. Most theoretical studies assume ideal continuously tunable RIS elements without mutual coupling (MC). Yet, the programmability of most existing RIS prototypes is constrained to a few (usually two) discrete states per RIS element due to the complexity of realizing circuits that can provide many independently and continuously controllable bias voltages~\cite{del2019optimally}. Moreover, MC effects can be significant~\cite{rabault2024tacit}. Strong MC implies a highly non-linear mapping from the RIS configuration to the corresponding end-to-end wireless channel~\cite{rabault2024tacit}. 

An MC-aware RIS optimization problem is generally high-dimensional, non-convex, and non-linear; hardware constraints like 1-bit tunable RIS elements further complicate the optimization. Closed-form solutions only exist for RIS with tunable inter-element coupling and without any hardware constraints~\cite{nerini2024global}.
Recent theoretical studies on MC-aware RIS optimization are model-based (i.e., using a physics-consistent system model whose parameters are assumed to be known) and assume continuously tunable RIS elements.
Incremental RIS configuration updates are at the core of these methods, either because utilized Neumann-series approximations inherently require sufficiently small configuration updates~\cite{qian2021mutual,abrardo2021mimo,ma2023ris,li2024beyond,abrardo2024design}, or because techniques like gradient-descent are used~\cite{el2023optimization,wijekoon2024phase,peng2025risnet}.
Therefore, none of these approaches can be straightforwardly applied to an RIS with 1-bit-programmable elements. Moreover, many of these theoretical studies simplify physics-consistent models from multi-port network theory (MNT) by assuming unilateral propagation~\cite{ivrlavc2010toward}, neglecting the MC between antennas and/or the MC between RIS elements originating from multi-bounce paths in the radio environment.

Meanwhile, a few experimental and full-wave numerical studies on realistic RIS hardware coped with 1-bit programmability and MC by using model-agnostic coordinate-descent (CD) algorithms~\cite{del2019optimally,tapie2024systematic,del2025physics}. However, these works did not systematically explore different optimization strategies, nor their dependence on the MC strength or potential benefits of intelligently initializing the optimization. 

Wireless practitioners urgently need guidance on which algorithm to choose to work with current RIS prototypes that are subject to MC and 1-bit-programmability; yet, to our knowledge, no systematic study comparing possible algorithms exists for discrete, MC-aware RIS optimization.

Moreover, existing theoretical works on physics-consistent RIS optimization are fundamentally limited by the difficulty of generating \textit{deterministic} channel realizations. On the one hand, theoretical studies require analytical tractability such that they 
can only explore very simple free-space scenarios without structural scattering~\cite{qian2021mutual}, sometimes adding fading effects ad hoc without considering their contribution to MC. On the other hand, end-to-end channel evaluations are costly in full-wave numerical studies of more complex scenarios; estimating model parameters of more complex scenarios is also quite costly in full-wave simulations~\cite{tapie2024systematic,zheng2024mutual} or experiments~\cite{sol2023experimentally,del2024virtual,del2025physics,del2025virtual,del2025experimental,del2025ambiguity}. A method to generate \textit{statistical} ensembles of physical-model parameters, ideally with controllable MC strength, is missing.
A statistical analysis for rich-scattering environments is desirable because MC generally depends on the specific environment~\cite{rabault2024tacit}; ensemble-averaged analysis provides robust, generalizable insights beyond any single realization.

Our contributions are summarized as follows. 
\textit{First}, we propose a simple reverberation-chamber-inspired statistical generator for MNT parameters of rich-scattering, RIS-parametrized radio environments, with a single hyper-parameter to
adjust the MC strength between RIS elements.
\textit{Second}, using this technique, we systematically compare model-agnostic and model-based strategies to optimize 1-bit-programmable RISs with different levels of MC strength in rich-scattering environments; we evaluate performance, execution time and memory usage. Where appropriate, we explore benefits of intelligently initializing the methods based on a dictionary search or a simplified affine model.

\textit{Notation:} $\mathbf{A}_\mathcal{BC}$ denotes the block of the matrix $\mathbf{A}$ whose row [column] indices are in the set $\mathcal{B}$ [$\mathcal{C}$]. $\mathrm{diag}(\mathbf{a})$ denotes the matrix whose diagonal entries are those of the vector $\mathbf{a}$ and all other entries are zero. $\mathbf{I}_a$ denotes the $a \times a$ identity matrix. 

\section{System Model}
\label{sec_syst_model}
In this section, we first describe the mathematical structure of the well-established MNT system model for RIS-parametrized radio environments and relate it to simplified models; then, we explain our simple technique for generating statistical ensembles of the MNT model parameters with controllable MC strength.

\begin{figure}
\centering
\includegraphics[width=0.7\columnwidth]{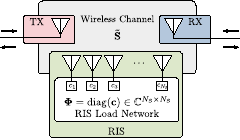}
\caption{MNT system model of an RIS-parametrized radio environment.}
\label{Fig1}
\end{figure}

\subsubsection{High-Fidelity Physics-Consistent MNT Model}

To apply MNT as shown in Fig.~\ref{Fig1}, we assume that the antenna ports are lumped and that the RIS elements' programmability originates from tunable lumped elements (e.g., PIN diodes). A tunable lumped element can be described by an auxiliary lumped port terminated by a tunable load.
With $N_\mathrm{T}$ transmitting antennas, $N_\mathrm{R}$ receiving antennas, and $N_\mathrm{S}$ RIS elements, the static parts of the RIS-parametrized radio environment are compactly characterized as a time-invariant, linear, passive, reciprocal $N$-port network, where $N=N_\mathrm{T}+N_\mathrm{R}+N_\mathrm{S}$. The corresponding scattering matrix $\tilde{\mathbf{S}}\in\mathbb{C}^{N\times N}$ incorporates the effects of all scattering objects in the radio environment (including structural scattering of the antennas and RIS elements)~\cite{prod2023efficient,sol2023experimentally,tapie2024systematic}. For simplicity, we use the same reference impedance $Z_0$ at all ports to define $\tilde{\mathbf{S}}$, and we assume that the signal generators and detectors attached to the transmitting and receiving antennas, respectively, are matched to $Z_0$. 

The end-to-end wireless channel matrix $\mathbf{H}\in\mathbb{C}^{N_\mathrm{R} \times N_\mathrm{T}}$ can be derived from MNT~\cite{li2024beyond,abrardo2024design,nerini2024universal}, yielding
\begin{equation}
    \mathbf{H} = \tilde{\mathbf{S}}_\mathcal{RT} +\tilde{\mathbf{S}}_\mathcal{RS} \left(\mathbf{\Phi}^{-1} - \tilde{\mathbf{S}}_\mathcal{SS}  \right)^{-1} \tilde{\mathbf{S}}_\mathcal{ST},
    \label{eq1}
\end{equation}
where $\mathcal{R}$, $\mathcal{T}$, and $\mathcal{S}$ denote the sets of port indices associated with receiving antennas, transmitting antennas, and RIS elements, respectively; $\mathbf{\Phi}=\mathrm{diag} (\mathbf{c})  \in \mathbb{C}^{N_\mathrm{S}\times N_\mathrm{S}}$, where $\mathbf{c}=[c_1, c_2, \dots, c_{N_\mathrm{S}}] \in \mathbb{C}^{N_\mathrm{S}}$ characterizes the reconfigurable parts of the RIS-parametrized radio environment (i.e., the RIS elements' tunable lumped components), $c_i$ being the reflection coefficient of the tunable load terminating the auxiliary port associated with the $i$th RIS element. 

Mathematically, the coupling of the influence of the different RIS elements on the wireless channel is mediated by the off-diagonal entries of $\tilde{\mathbf{S}}_\mathcal{SS}$. This directly correlates with the ease of inverting  $(\mathbf{\Phi}^{-1} - \tilde{\mathbf{S}}_\mathcal{SS})$ in (\ref{eq1}). In the limit of vanishing off-diagonal entries of $\tilde{\mathbf{S}}_\mathcal{SS}$, the coupling vanishes and the inversion of $(\mathbf{\Phi}^{-1} - \tilde{\mathbf{S}}_\mathcal{SS})$ is trivial. 
Following~\cite{prod2025benefits}, we thus quantify the MC strength between RIS elements as
\begin{equation}
    \mu_\mathrm{n} = \left\langle \frac{\left\lVert \tilde{\mathbf{S}}_\mathcal{SS} - \mathrm{diag}\left( \tilde{\mathbf{S}}_\mathcal{SS} \right) \right\rVert_2}{ \left\lVert \mathbf{\Phi}^{-1} - \mathrm{diag}\left( \tilde{\mathbf{S}}_\mathcal{SS} \right) \right\rVert_2} \right\rangle,
\end{equation}
where $\langle \cdot \rangle$ denotes the ensemble averaging over random RIS configurations so that we have a configuration-agnostic metric.\footnote{We average over 100 random RIS configurations to ensure a small standard error of the mean  while keeping runtime reasonable.}

\subsubsection{Low-Fidelity Affine Models}

To relate the MNT model to widespread simplified models, it is instructive to rewrite~(\ref{eq1}) as infinite series~\cite{del2025physics,zheng2024mutual,wijekoon2024phase}:
\begin{equation}
    \mathbf{H} = \tilde{\mathbf{S}}_\mathcal{RT} +\tilde{\mathbf{S}}_\mathcal{RS} \left[\sum_{k=0}^{\infty}(\mathbf{\Phi}\tilde{\mathbf{S}}_\mathcal{{SS}})^k\right] \mathbf{\Phi} \tilde{\mathbf{S}}_\mathcal{ST},
    \label{eq2}
\end{equation}
where passivity guarantees convergence~\cite{del2025physics}. The $k$th term of the series captures all paths involving $k$ bounces between RIS elements~\cite{rabault2024tacit,del2025physics}. The spectral radius $R$ (i.e., the largest eigenvalue magnitude) of $\mathbf{\Phi}\tilde{\mathbf{S}}_\mathcal{{SS}}$ governs the convergence speed of the series in (\ref{eq2}). The series collapses to a single term only if all entries of $\tilde{\mathbf{S}}_\mathcal{{SS}}$ vanish, in which case one recovers the widespread simplified cascaded (CASC) model with an affine mapping 
\begin{equation}
  \mathbf{H}_\mathrm{casc} = \tilde{\mathbf{S}}_\mathcal{RT} +\tilde{\mathbf{S}}_\mathcal{RS} \mathbf{\Phi} \tilde{\mathbf{S}}_\mathcal{ST}
  \label{eq3}
\end{equation}
from $\mathbf{\Phi}$ to $\mathbf{H}$. Importantly, zero MC alone does not guarantee this collapse of the series. Indeed, if the diagonal entries of $\tilde{\mathbf{S}}_\mathcal{{SS}}$ (i.e., the reflection coefficients seen at the auxiliary ports with all other ports matched to $Z_0$) do not vanish, the series does not collapse. Physically, the series represents bounces confined to the interfaces between radio environment and load network in such cases, i.e., \textit{not} between different auxiliary lumped ports. These bounces at the $i$th auxiliary port arise  because the radio environment does not present an impedance of $Z_0$ to the $i$th load. Hence, the value of $R$ is generally not directly indicative of the MC strength.

A completely unstructured, affine model, referred to as RR in the following, can be obtained via ridge regression based on $M$ pairs of $\mathbf{H}$ and $\mathbf{\Phi}$, each pair corresponding to a distinct random RIS configuration.

\subsubsection{Statistical MNT Model Parameter Generation}

Having described and analyzed the mathematical structure of the MNT system model, we now discuss the choice of MNT-model parameters which are the allowed values of $c_i$ and the entries of $\tilde{\mathbf{S}}$. \textit{Regarding the former}, most theoretical studies assume that $c_i$ can take any complex value whose magnitude does not exceed unity (due to passivity), or that $c_i$ can be continuously tuned based on a model that intertwines phase and amplitude response~\cite{abeywickrama2020intelligent}. In contrast, given our focus on 1-bit programmability, we work with $c_i\in\{ -1,1\}$ in this Letter. \textit{Regarding the latter}, we now introduce a simple technique for generating statistical ensembles of $\tilde{\mathbf{S}}$ that emulate rich-scattering environments with tunable MC between RIS elements. This is in contrast to deterministic approaches that evaluate $\tilde{\mathbf{S}}$ for a concrete setup, e.g., using a discrete-dipole approximation~\cite{PhysFad,mursia2023saris}, for which the generation of large ensembles of $\tilde{\mathbf{S}}$ is prohibitively costly.

Our technique builds upon the statistical description of scattering in reverberation chambers (RCs)~\cite{holloway2006use,garcia2011advances,chen2018reverberation}.
In an ideal mode-stirred RC with at least half-wavelength antenna spacing, the channel coefficients between antenna ports (which are off-diagonal entries of a scattering matrix) are drawn from a circularly symmetric zero-mean complex Gaussian distribution (Rayleigh fading).
Moreover, in this setting the variance of the reflection coefficients is twice that of the transmission coefficients due to coherent backscattering~\cite{ladbury2007enhanced}. 
To map these insights to our system model, we recall that RIS elements are antennas terminated by tunable loads. Thus, $\tilde{\mathbf{S}}$ corresponds to the scattering matrix of an ideal reverberation chamber containing $N$ antennas with at least half-wavelength spacing. The half-wavelength spacing between adjacent RIS elements is common in most RIS prototypes~\cite{del2019optimally}. We hence draw the diagonal and off-diagonal entries of $\tilde{\mathbf{S}}$ from circularly symmetric zero-mean complex Gaussian distributions with variance 1 and 0.5, respectively. We impose that $\tilde{\mathbf{S}}$ is symmetric to respect reciprocity. 
To respect passivity, we scale all entries of $\tilde{\mathbf{S}}$ by a factor $\alpha$ and subsequently discard any realizations of $\tilde{\mathbf{S}}$ that have singular values above unity; we heuristically chose $\alpha=1/15$ but we checked that other choices of $\alpha$ that ensure passivity lead to the same qualitative behavior of our results. 

Finally, we introduce \textit{ad hoc} a simple single-parameter tuning knob to adjust the RIS elements' MC strength: a positive, real hyper-parameter $\kappa$ scales all off-diagonal entries of $\tilde{\mathbf{S}}_\mathcal{SS}$. The value of $\mu_\mathrm{n}$ scales by definition with the same factor $\kappa$; we impose an upper limit on $\kappa$ to respect the passivity of $\tilde{\mathbf{S}}$. 
The simplicity of this \textit{ad hoc} MC strength control, not requiring the incorporation of detailed geometrical information, justifies its use to study algorithmic performance without committing to any particular hardware layout.

\section{MC and Binary-Tunability Aware Optimization}
\label{sec_algs}
This section introduces the optimization algorithms that we consider to minimize a cost $\mathcal{C}(\mathbf{H})$;  we define $\mathcal{C}(\mathbf{H})$ as the negative of the considered performance indicator in Sec.~\ref{sec_perf_eval}.

\subsubsection{Dictionary Search (DS)}
DS is a brute-force search over a dictionary of $M$ pairs of a randomly generated RIS configuration and the corresponding channel matrix, evaluated using the considered model. The lowest-cost configuration is selected. We only explore DS based on the MNT model.

\subsubsection{Coordinate Descent (CD)}
Starting with an initial RIS configuration $\mathbf{c}_\mathrm{init}$, the CD Algorithm~\ref{alg:cd} loops over each RIS element in turn to check based on the considered model whether changing its state lowers the cost. It stops after an entire loop without updating the RIS configuration. 
We consider CD based on all three models introduced in Sec.~\ref{sec_syst_model}. 
For RR-CD, we consider random initialization; for CASC-CD and MNT-CD, we consider initializations based on the outcome of DS or RR-CD. Both initializations depend on a dictionary of $M$ matching pairs of $\mathbf{H}$ and $\mathbf{\Phi}$, such that $M=0$ corresponds to random initialization.
We efficiently implement the required forward evaluations of the MNT model after a single-entry update of $\mathbf{c}$ using the Woodbury identity~\cite{prod2023efficient}.

\begin{algorithm}
\begingroup\small
\caption{{\small Binary Coordinate Descent}}
\label{alg:cd}
Initialize RIS configuration: \( \mathbf{c}_{\mathrm{curr}} \gets \mathbf{c}_{\mathrm{init}} \). \\
$\mathcal{C}_\mathrm{curr} \gets \mathcal{C}(\mathbf{H}(\mathbf{c}_\mathrm{curr}))$. \\
$k \gets 0$, $j \gets 0$.\\
\While{\( k < N_{\mathrm{S}} \)}{
        $j \gets (j \ \mathrm{mod} \ N_\mathrm{S}) + 1$\\
        \(\mathbf{c}_\mathrm{new} \gets \mathbf{c}_\mathrm{curr}\) with its $j$th entry flipped.\\
        $\mathcal{C}_\mathrm{new} \gets \mathcal{C}(\mathbf{H}(\mathbf{c}_\mathrm{new}))$.  \\
        \eIf{\( \mathcal{C}_\mathrm{new} < \mathcal{C}_\mathrm{curr} \)}{
         \( \mathbf{c}_{\mathrm{curr}} \gets \mathbf{c}_{\mathrm{new}} \); 
         \( \mathcal{C}_{\mathrm{curr}} \gets \mathcal{C}_{\mathrm{new}} \); 
            \( k \gets 0 \).
        }
        {
            \( k \gets k+1 \).
        }
    
}
\textbf{Output:} Optimized RIS configuration \( \mathbf{c}_{\mathrm{curr}}. \)
\endgroup
\end{algorithm}

\subsubsection{Temperature-Annealed Back-Propagation (TABP)}
TABP optimizes a relaxed continuous version of the binary problem using gradient descent, gradually driving the distribution toward a binary one~\cite{chakrabarti2016learning,del2020learned}. Specifically, a continuous, real-valued variable $\mathbf{z}\in \mathbb{R}^{N_\mathrm{S}}$ is mapped to $\tilde{\mathbf{c}}$ using the temperature-scaled \( \tanh \) function: $ \tilde{\mathbf{c}}(\mathbf{z},t) = c' + \frac12\bigl(1 + \tanh(\frac{\mathbf{z}}{t})\bigr)(c''-c')$, where $c'$ and $c''$ are the two allowed values for the entries of $\mathbf{c}$, and $t$ is the temperature parameter. 
As the epochs advance, $t$ decreases and the distribution of the entries of $\tilde{\mathbf{c}}$ becomes more and more binary. After training, we enforce the binarization: $\mathbf{c}_{\mathrm{curr}} = \lim_{t \to 0} \tilde{\mathbf{c}}(\mathbf{z},t)$.
We implement the TABP Algorithm~\ref{alg:bp_temp} based on the CASC and MNT models, using the Adam optimizer~\cite{kingma2014adam} with default hyper-parameters to minimize $\mathcal{C}(\mathbf{H})$; we consider the same initializations as for CASC-CD and MNT-CD. We train for up to $e_\mathrm{max}=200$ epochs with $\epsilon = 10^{-6}$, $t_\mathrm{start}=1$, and $t_\mathrm{end}=0.1$.

\begin{algorithm}
\begingroup\small
\caption{{\small Temperature-Annealed Back-Propagation}}
\label{alg:bp_temp}
\(\displaystyle \mathbf{z} \;\gets\; t_{\rm start}\,\atanh ((1-\epsilon)(2\tfrac{\mathbf{c}_{\rm init}-c'}{c''-c'} - 1))\). \\
$\tilde{\mathbf c}\gets c' + \tfrac12\bigl(1+\tanh\!\bigl(\tfrac{\mathbf z}{t_{\rm start}}\bigr)\bigr)(c''-c')$.\\
$\mathcal{C}_\mathrm{curr} \gets \mathcal{C}(\mathbf{H}(\tilde{\mathbf{c}}))$.\\
\For{\(e=1\) to \(e_{\max}-1\)}{
    $t \gets t_{\text{start}} \left( \tfrac{t_{\text{end}}}{t_{\text{start}}} \right)^{e/e_\mathrm{max}}$.\\
    $\tilde{\mathbf{c}} \gets c' + \frac12\bigl(1 + \tanh(\frac{\mathbf{z}}{t})\bigr)(c''-c')$.\\
    $\mathcal{C}_\mathrm{new}\gets\mathcal{C}(\mathbf{H}(\tilde{\mathbf{c}}))$.\\
    \If{$\left| \mathcal{C}_\mathrm{new} - \mathcal{C}_\mathrm{curr} \right| \leq \epsilon$}{
        \textbf{break}\\
    }
    $\mathbf{g} \gets \nabla_\mathbf{z} \mathcal{C}_\mathrm{new}$.\\
    $\mathbf{z} \gets \texttt{AdamStep}(\mathbf{z}, \mathbf{g})$.\\
    $\mathcal{C}_{\text{curr}} \gets \mathcal{C}_\mathrm{new}$.\\
}
$\mathbf{c}_{\mathrm{curr}} = \lim_{t \to 0} \tilde{\mathbf{c}}(\mathbf{z},t)$.\\
\textbf{Output:} Optimized RIS configuration \( \mathbf{c}_{\mathrm{curr}}. \)
\endgroup
\end{algorithm}

\subsubsection{Genetic Algorithm (GA)}
Inspired by natural evolution~\cite{lambora2019genetic}, our GA maintains a population of \(M\) candidate RIS configurations and iterates for 10 generations. At each generation, we compute each individual’s fitness $F = -\mathcal{C}(\mathbf{H})$ and select the top \(M/2\) to form a mating pool. We then generate \(M\) offspring by repeatedly sampling two parents (with replacement), performing one-point crossover, and mutating each child with probability \(10^{-4}\).  We only consider GA based on the MNT model.

Of the considered algorithms, only TABP is inevitably model-based because of the required gradient evaluations. All other algorithms can be implemented in a model-agnostic manner, e.g., based on experimental measurements of $\mathbf{H}(\mathbf{c})$ without disposing of a calibrated model, as in~\cite{del2019optimally}.

\section{Performance Evaluation}
\label{sec_perf_eval}

We consider a prototypical single-input single-output (SISO) channel-gain optimization problem with a 100-element RIS. Specifically, we define $\mathcal{C}(\mathbf{H}) = -|S_{21}|^2$. Despite being an antenna-level metric, the SISO channel gain is a commonly considered performance indicator in theoretical papers on RIS optimization~\cite{qian2021mutual,li2024beyond} because it directly relates to end-to-end SISO communications metrics like capacity and bit-error rate. We defer the more complicated analysis of multiple-antenna cases to future work. 
Besides the  channel gain achieved with the algorithms introduced in Sec.~\ref{sec_algs}, we also evaluate the corresponding execution time and memory usage.

Because CD and TABP require an initialization, they are amenable to multi-fidelity optimization. We examine the benefit of initializing them with the lowest-cost configuration found with DS or RR-based CD given a dictionary of $M$ random RIS configurations and the corresponding MNT-based end-to-end channels. We systematically study the influence of the choice of $M$ ($M=0$ corresponds to random initialization), as well as of the MC strength $\mu_\mathrm{n}$ and the assumed model (MNT or CASC). To achieve statistical robustness, we repeat all optimizations for 1500 independent realizations of \( \tilde{\mathbf{S}}\).

\begin{figure*}
\centering
\includegraphics[width=\textwidth]{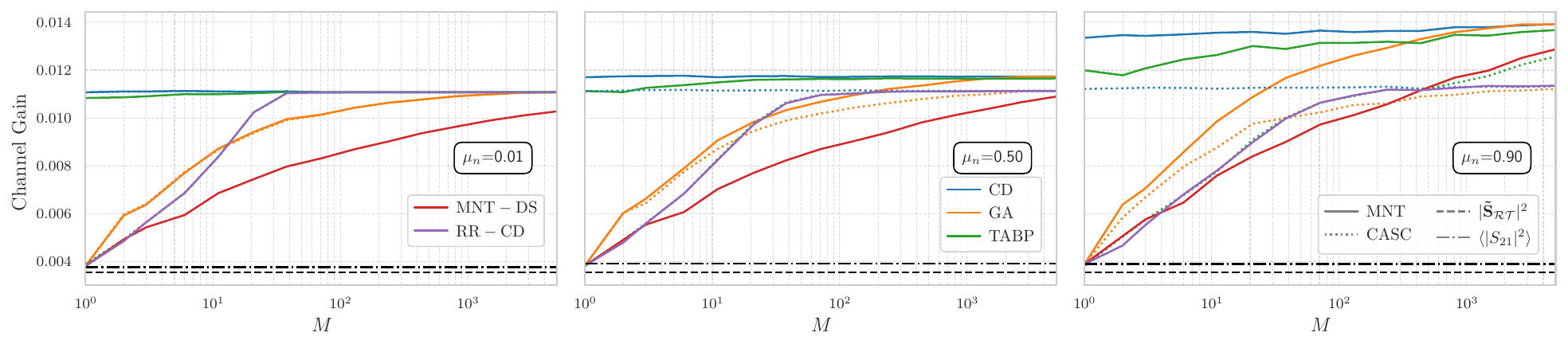}
\caption{SISO channel gain $|S_{21}|^2$ for the considered methods as a function of $M$ (horizontal axis) and $\mu_\mathrm{n}$ (panels), averaged over 1500 realizations.}
\label{Fig2}
\end{figure*}

\textit{Achieved Channel Gains:}
Our results displayed in Fig.~\ref{Fig2} reveal that in the regime of weak MC (left panel; \( \mu_\mathrm{n} = 0.01 \)), all algorithms achieve nearly identical channel gains (for sufficiently high $M$ in some cases), even RR-based CD for $M>N_\mathrm{S}$. 
Hence, as one may expect, the affine approximation is very accurate in the weak-MC regime; indeed, the curves for CASC (dashed) and MNT (continuous) are superposed in this case. As the MC level is increased, the approximation underlying CASC naturally becomes less accurate and the differences between CASC and MNT become apparent (middle and right panel). In the regime of moderate MC (middle panel; $\mu_\mathrm{n}=0.5$) the performance gap is already notable, and in the regime of strong MC (right panel; $\mu_\mathrm{n}=0.99$) it is large. 
Across all regimes, we observe that MNT-based CD achieves the largest channel gains without  significant dependence on $M$. Thus, no benefits of intelligently initializing CD are apparent. The channel gains of TABP and GA are almost on a par with those of CD for large $M$.

\begin{figure}[h]
\centering
\includegraphics[width=0.85\columnwidth]{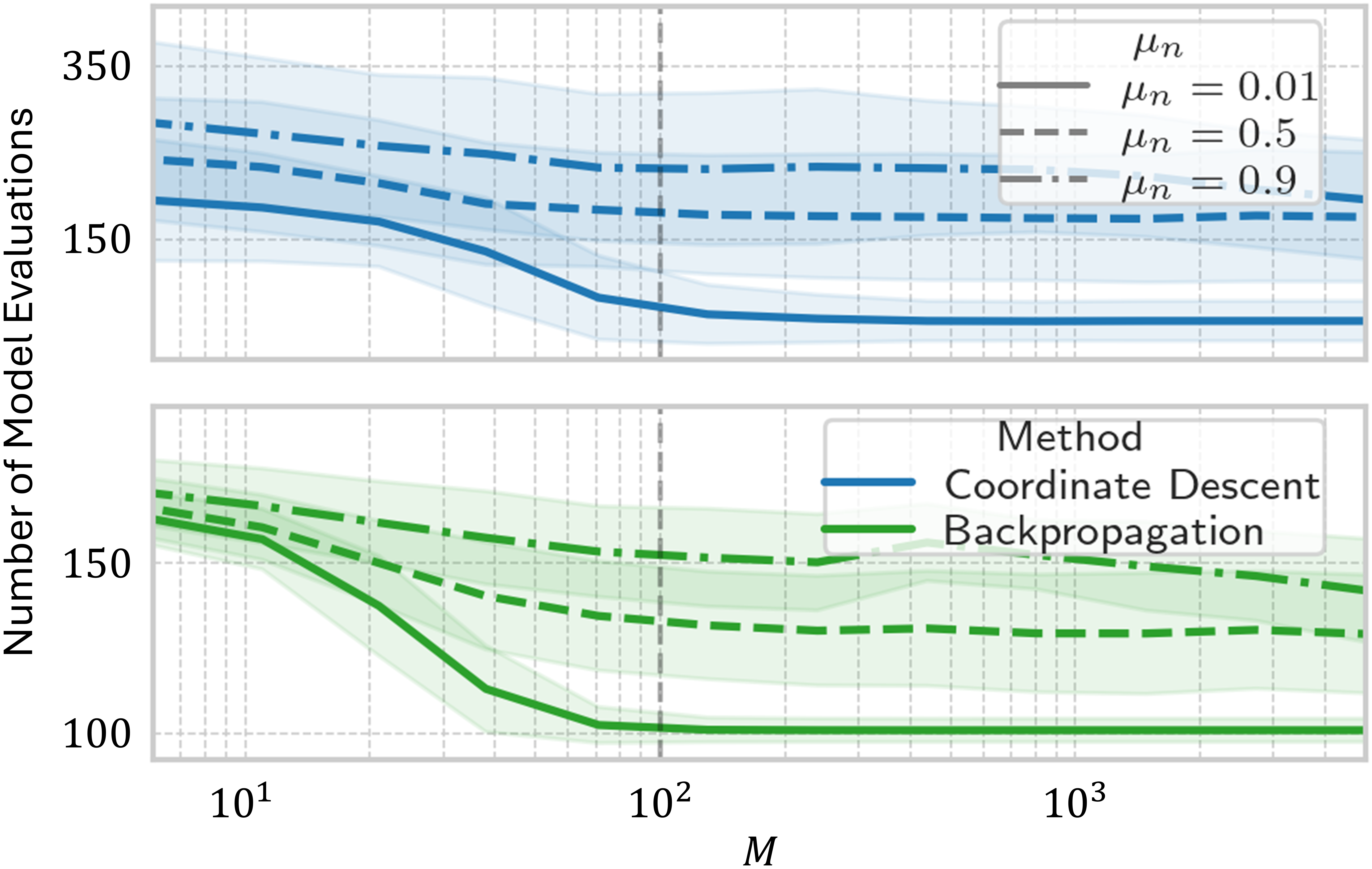}
\caption{Impact of $M$ and $\mu_\mathrm{n}$ on the methods' convergence.}
\label{Fig3}
\end{figure}

\textit{Computational Complexity}:
We assess the temporal complexity of each optimization algorithm based on the number of evaluations of the model (irrespective of the model's fidelity) required for convergence or to reach a fixed iteration limit. The number of evaluations equals the dictionary size $M$ for DS, and $10M$ for GA with 10 generations of population size $M$. TABP and CD typically require \(\mathcal{O}(N_\mathrm{S})\) model evaluations, as seen in Fig.~\ref{Fig3}. We also observe that better initializations (based on larger $M$) reduce the temporal complexity of TABP and CD by accelerating their convergence, especially in regimes with lower MC strength. For strong MC, the benefits of a good initialization in terms of the convergence speed are small; in fact, the reduction in the number of model evaluations is smaller than $M$ in this case. We also observe that generally the benefits of initialization for the convergence speed saturate when $M>N_\mathrm{S}$. Finally, we see in Fig.~\ref{Fig3} that TABP and CD generally converge faster for lower MC strengths.

While the number of model evaluations is somewhat smaller for TABP than for CD, we note that, \textit{first}, the Woodbury identity enables efficient computations of the model evaluations in CD but not in TABP~\cite{prod2023efficient}, and, \textit{second}, TABP additionally requires gradient computations that are not accounted for in our temporal complexity metric. 
Unlike CD and TABP, DS and GA are amenable to parallelization, which can substantially shorten the optimization times provided suitable computational resources.

\textit{Memory Complexity:}
To characterize the algorithms' memory usage, we estimate the size of the variables stored at each iteration. The memory usage of CD and TABP is \(\mathcal{O}(N_\mathrm{S})\): it only depends on $N_\mathrm{S}$ because they only store the current configuration (and gradients for TABP). In contrast, DS, the ridge regression underlying RR-CD, and GA require storing a large set of configurations (for evaluation, surrogate model training, or population evolution); their memory usage grows linearly with the number of configurations, as \(\mathcal{O}(MN_\mathrm{S})\).

\section{Conclusion}

To summarize, we introduced a simple statistical approach to MNT-based modeling of RIS-parametrized channels and leveraged it to comprehensively compare optimization strategies for RIS with MC and limited to 1-bit-programmable elements, relevant to most existing RIS prototypes.
The choice of optimization method should be guided by the MC strength and the available computational resources.  
For weak MC, simple methods like DS or RR-CD are sufficient and lightweight. 
For higher MC, CD and TABP are more effective, with CD offering the best trade-off between performance, speed, and memory. Both methods achieve comparable channel gains and require a similar number of model evaluations; however, TABP requires additional gradient computations, cannot benefit from the Woodbury identity, and inevitably requires a calibrated model. The cost of intelligent initialization could not be amortized, especially not for strong MC. GA can be competitive under parallel execution with loose resource constraints. 

Altogether, we conclude that for RIS with non-negligible MC and 1-bit-programmable elements, CD offers the best compromise between channel gain, computational efficiency, and memory usage.
We expect our insights to generalize to other reconfigurable wave systems (e.g., dynamic metasurface antennas and wave-domain physical neural networks).

Looking forward, we envision an extension of our statistical MNT-based model-parameter generation to broadband scenarios. Therein, spectral correlations must be captured correctly, which can be achieved with the so-called Heidelberg random-matrix-theory approach~\cite{beenakker1997random,viviescas2003field}. This approach can be of interest even in the single-frequency case to capture higher-order correlations between the entries of $\tilde{\mathbf{S}}$; however, Rayleigh fading (only concerned with $\mathbf{H}$ in the absence of an RIS) ignores such correlations.  Moreover,  the presented analysis can be extended to account for overhead associated with estimating the model parameters.

\bibliographystyle{IEEEtran}

\end{document}